\documentclass[aps,prl, eprint, twocolumn, amsmath, bibliography]{revtex4-1}
\usepackage{amsmath}
\usepackage{graphicx}
\usepackage{dcolumn}
\usepackage{natbib}
\usepackage{float}
\begin{document}
\title{Three-dimensional imaging of a pattern localized in a phase space}
\author{Mandip Singh and Samridhi Gambhir}
%\email{mandip@iisermohali.ac.in}
\affiliation{Department of Physical Sciences,
Indian Institute of Science Education and Research Mohali, Mohali, 140306, India.}
\begin{abstract}
In coventional imaging experiments, objects are localized in a position space and such optically responsive objects can be imaged with a convex lens and can be seen by a human eye.  In this paper, we introduce an experiment on a three-dimensional imaging of a pattern which is localized in a three-dimesional phase space. The phase space pattern can not be imaged with a lens in a conventional way and it can not be seen by a human eye. In this experiment, a phase space pattern is produced from object transparancies and imprinted onto the phase space of an atomic gaseous medium, of doppler broadened absorption profile at room temperature, by utilizing velocity selective hole burning in the absorption profile. The pattern is localized in an unique three dimensional phase space which is a subspace of the six dimensional phase space.  Imaging of the localized phase space pattern is performed at different momentum locations. In addition, imaging of the imprinted pattern of an object of nonuniform transmittance is presented. 
 
\end{abstract}
\maketitle

%\subsection{Introduction}
In most imaging experiments, a structure of an object is defined in a position space. The structural pattern can be stationary or for a dynamic object can be non stationary \emph{w.r.t.} time. An image of such an optically responsive object can be produced with a convex lens therefore, such an object can be seen with a camera or by a human eye. In this paper, we go beyond the conventional notion of imaging. A structural pattern of objects in our experiment is defined in a phase space therefore, such a pattern can not be imaged with a lens or a camera and a human eye can not visualize such a pattern.
In this paper, we introduce a three-dimensional (3D) imaging of a pattern localized in a phase space. The pattern is localized in an unique 3D subspace, of the six-dimensional (6D) phase space, involving two position and one momentum coordinates. However, the pattern is delocalized in a 3D position subspace and in a 3D momentum subspace of the 6D phase space, separately.   

In experiment presented in this paper, the pattern of interest is produced by object transparencies and imprinted onto the phase space of an atomic gaseous medium at room temperature. Experiment is performed by utilizing velocity selective hole-burning \cite{lamb, bennet, haroche, hughes2, scholten1, boudot, schm} in doppler broadened absorption profile of an atomic gaseous medium. Tomographic images of the pattern localized in a 3D phase space are then captured with an imaging laser beam. The imaging laser beam is not interacting with actual objects used to produce the localized phase space pattern. Imaging of objects localized in a position space has been realized with quantum $\&$ classical methods with undetected photons \cite{zeilinger_1, wong}. Quantum imaging with undetected photons, and unlike the ghost imaging \cite{bar, imphase, qimaging, ghim, boyd, shih, lugiato}, does not rely on coincidence detection of photons. In this paper,  a pattern of an object of nonuniform transmittance is also imprinted onto the phase space of an  atomic medium and the pattern is then imaged at a constant location of momentum.

\begin{figure}[H]
\begin{center}
\includegraphics[scale=0.29]{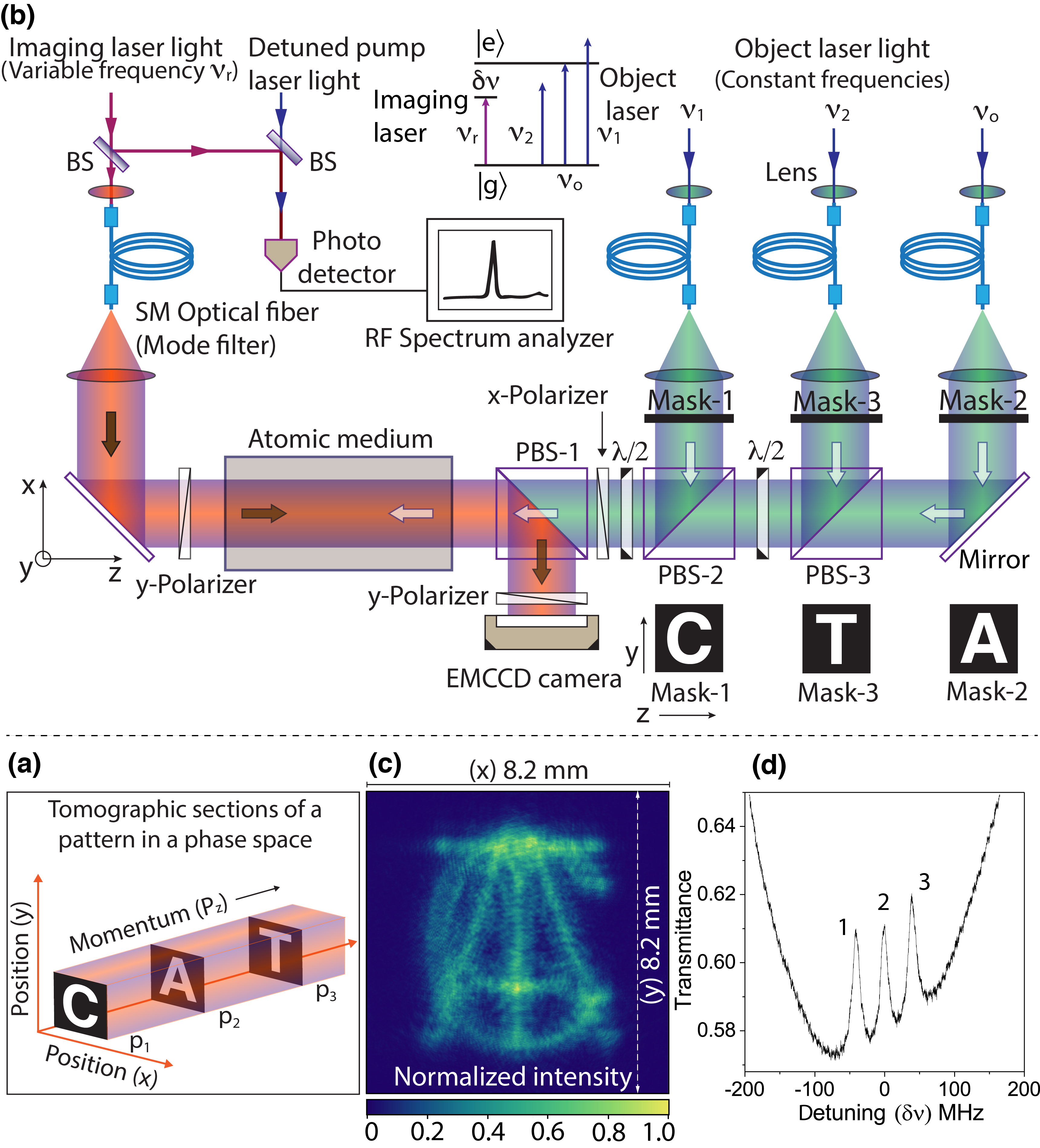}
\caption{\label{fig1} \emph{(a)  A localized pattern in a 3D phase space and its three tomograms. (b) Experimental schematic diagram, a linearly polarized imaging laser beam is overlapped in an atomic gaseous medium with counter propagating object laser beams. A 2D transverse intensity profile of the imaging laser beam at different detunings is captured with an EMCCD camera. (c) A 2D transverse intensity profile of the overlapped object laser beams prior to their entrance into the atomic medium.  All three alphabets are overlapped with each other. (d) Transmittance,  for imaging laser beam, of the atomic medium in presence of object laser beams without masks. Three peaks labeled as $1$, $2$ and $3$ correspond to a velocity selective hole-burning, in doppler broadened absorption profile, produced by object laser beams of frequencies $\nu_{1}$, $\nu_{o}$ and $\nu_{2}$, respectively.}}
\end{center}
\end{figure}

A localized pattern in a 3D subspace of the 6D phase space is shown in Fig.~\ref{fig1} (a), where a two dimensional position space is spanned by orthogonal position unit vectors $\hat{x}$ and $\hat{y}$ and a third dimension corresponds to the $z$-component of momentum, $p_{z}$. 

\vspace{5mm}
In experiment, $p_{z}$ is the $z$-component of momentum of atoms. The pattern is stationary \emph{w.r.t.} time.
 A 2D planar section of a localized 3D phase space pattern at a constant $p_{z}$ represents a tomogram of the localized pattern. In Fig.~\ref{fig1} (a), three different tomograms at three different momenta are shown. Tomograms with an image of the English script alphabets $\bf{C}$, $\bf{A}$ and $\bf{T}$ are localized at $p_{z}$ equals to $p_{1}$, $p_{2}$ and $p_{3}$, respectively. Furthermore,  in a 3D position space, spanned by orthogonal position unit vectors $\hat{x}$, $\hat{y}$ and $\hat{z}$, each tomogram is completely delocalized on $z$-axis that implies,  all images are overlapped with each other and distributed at all points on $z$-axis. 
However, in a 3D momentum space, spanned by orthogonal unit vectors $\hat{p}_{x}$, $\hat{p}_{y}$ and $\hat{p}_{z}$ of momentum components along $\hat{x}$, $\hat{y}$ and $\hat{z}$ directions, each tomogram is delocalized in all planes parallel to $p_{x}$-$p_{y}$ plane.
 A subspace where the pattern is completely localized is an unique 3D subspace of the 6D phase space, as shown in Fig.~\ref{fig1} (a). In remining 3D subspaces of the 6D phase space the pattern is delocalized. In this paper, stationary localized 3D phase space pattern of interest is produced from objects located in the position space. The pattern is then imprinted onto the phase space of an atomic gas obeying Maxwell velocity distribution, in form of difference of number density of atoms in ground state and excited state.  Tomographic images of the 3D phase space pattern are then imaged with an imaging laser beam, where by varing the frequency of the laser beam the location, $p_{z}$, of the tomogram can be shifted. 

In experiment, a stationary pattern in phase space of atoms is produced at room temperature (25$^{o}$C) by velocity selective hole-burning in the doppler broadened absorption profile of an atomic gaseous medium. Consider a linearly polarized object laser beam, of frequency $\nu_{p}$ and transverse intensity profile $I_{p}(x,y,\nu_{p})$, propagating in an atomic gaseous medium in a direction opposite to $z$-axis. The intensity profile $I_{p}(x,y,\nu_{p})$ represents a 2D image of an object in position space.  This image information is transferred to a velocity class of the atomic gaseous medium at temperature $T$ by velocity selective atomic excitation. Consider an atomic gaseous medium where an isolated stationary atom has a ground quantum state $|g\rangle$ of energy $E_{g}$ and an excited quantum state $|e\rangle$ of energy $E_{e}$ with linewidth $\Gamma$. The object laser beam is on resonance with a velocity class of atoms of $z$-component of their velocity equals to $v_{r}=2\pi(\nu_{o}-\nu_{p})/k$, where  $\nu_{o}=(E_{e}-E_{g})/h$, $k=2\pi/\lambda$ is the magnitude of the propagation vector of the object laser beam having wavelength $\lambda$.  Atoms of other velocity classes are out of resonance due to the doppler shift. Transverse doppler shift is negligible beacuse of non relativistic velocity regime at room temperature. In absence of an object laser beam, all the atoms are in the ground state. Consider $n$ is the number of atoms per unit volume of the gaseous medium. According to Maxwell velocity distribution, a fraction of atoms with velocity in an interval $dv_{z}$ around $v_{z}$ at temperature $T$ is $f(v_{z}) dv_{z}=(m/2 \pi k_{B} T)^{1/2}e^{-m v^{2}_{z}/2 k_{B} T} dv_{z}$. Where, $k_{B}$ is the Boltzmann constant and $m$ is mass of an atom. Consider $L$ is length of the atomic medium along beam propagation direction. In presence of an object laser beam the ground state atoms of resonant velocity class are populated to the excited state. A steady state difference of atomic number densities in the ground state ($n_{1}$) and in the excited state ($n_{2}$) at $v_{z}$ is $n_{1}(x,y,v_{z})-n_{2}(x,y,v_{z})=n f(v_{z})/(1+I_{p}({x,y,\nu_{p}})\Gamma^{2}/(4I_{s}((2 \pi \nu_{p}-2 \pi \nu_{o}+kv_{z})^{2}+\Gamma^{2}/4)))$, where $I_{s}$ is the saturation intensity of the atomic transition. If attenuation and diffraction of the object laser beam are negligible then the transverse intensity profile of object laser beam is imprinted in the form of an atomic population difference in the resonating velocity class of atoms. This pattern is delocalized in the longitudinal direction \emph{i.e.} in the direction of propagation of the object laser beam. However, the pattern is localized in the transverse plane of coordinates $x$, $y$ at a $z$-component of momentum of atoms, $p_{z} = m v_{r}$. If three different overlapping object laser beams of same linear polarization, different intensity profiles and frequencies are passed through the atomic medium then each beam imprints a different pattern in a different velocity class. Where each one located at a different $p_{z}$ corresponding to the resonant velocity class of atoms adressed by the resonating object laser beam. As a result a localized patten of all objects is imprinted onto a 3D subspace of the 6D phase space of atoms.  The nearest frequency separation of object laser beams has to be much larger than the linewidth of the transition to reduce the overlapping of resonating velocity classes. 

To image the localized phase space pattern, a counter propagating imaging laser beam of frequency $\nu$ is overlapped with the object laser beams passing through the atomic gaseous medium. The polarization of the imaging laser beam is perpendicular to the polarization of object laser beams. The total absorption coefficient $\alpha$ of the imaging laser beam at frequency detuning, $\delta\nu=\nu-\nu_{o}$, is a convolution of population difference and absortion crossection of an atom such that $\alpha(x,y,\delta\nu) =\int^{\infty}_{-\infty} [n_{1}(x,y,v_{z})-n_{2}(x,y,v_{z})] \sigma_{o}(\Gamma^{2}/4) dv_{z}/((2 \pi \delta\nu- kv_{z})^{2}+\Gamma^{2}/4)$, where $\sigma_{o}$ is the peak absorption crossection of the atomic transition. The absorption of the imaging laser beam decreases if it interacts with a velocity class of atoms excited by object laser beams \emph{i.e.} $n_{2}(x,y,v_{z})$ is nonzero. This produces velocity selective hole-burning in the doppler broadened absorption profile of the atomic medium. For the incident transverse intensity profile of the imaging laser beam $I_{r}(x,y,\delta\nu)$, the transmitted imaging laser beam intensity profile after passing through the gaseous medium is     
$I_{r}(x,y,\delta\nu) \exp(-\mathrm{OD}(x,y,\delta\nu))$. Where $\mathrm{OD}(x,y,\delta\nu)=\alpha(x,y,\delta\nu) L$ is the optical density of the atomic medium. The optical density profile, at a detuning $\delta\nu$, corresponds a tomographic section of the phase space pattern at $p_{z}= 2 \pi m \delta\nu /k$. The optical denisty of the medium decreases if object laser beams are present. An image of a tomographic section can be constructed by measuring a change in the optical density profile caused by object laser beams. A 3D image of the phase space pattern can be constructed with tomograms obtained at different detunings of imaging laser beam. 

In experiment, objects are three 2D transparency masks where each mask consists of an image of an alphabet of the English script, 
$\bf{C}$ (on mask-$1$), $\bf{A}$ (on mask-$2$) and $\bf{T}$ (on mask-$3$), as shown in Fig.~\ref{fig1} (b). 
All alphabets are transparent and remaining part of each mask is completely opaque to light.  Object laser beams are initially passed through single mode (SM) polarization maintaining optical fibers to produce beams of gaussian transverse intensity profile. Where optical fibers are utilized as  transverse mode filters. The mode filtered and collimated object laser beams of frequencies $\nu_{1}$, $\nu_{o}$ and $\nu_{2}$ are then passed through the mask-$1$, mask-$2$ and mask-$3$, respectively.
After the masks, the transverse intensity profile of object laser beams  correspond to alphabets $\bf{C}$ (at $\nu_{1}$), $\bf{A}$ (at $\nu_{o}$) and $\bf{T}$ (at $\nu_{2}$). All three object laser beams are overlapped on polarization beam splitters (PBS-$3$, PBS-$2$). The overlapped object laser beams are linearly $x$-polarized by a polarizer with its pass axis aligned along $x$-axis ($x$-polariser). Two half wave-plates are placed, before and after the PBS-$2$, to rotate the linear polarization of the object laser beams to equialize the intensity. The image of transverse intensity profile of the overlapped object laser beams, prior to their entrance into the atomic medium, is shown in Fig.~\ref{fig1} (c) where, images of alphabets are overlapped with each other. A different alphabet is imprinted on a light field of different frequency and momentum.
Therefore, an intensity profile of each object laser beam also corresponds to a tomograph in the 3D phase space.

The overlapped object laser beams are passed through an atomic gaseous medium, which is a $10$~cm long rubidium ($^{87}$Rb)  vapour cell shielded from external magnetic field. The  linewidth of resonant transition of the atomic medium is broadened due to doppler shift caused by motion of atoms. An object laser beam of frequency $\nu_{o}$ is on resonance to the atomic transition of stationary  $^{87}$Rb atoms where a  ground quantum state is $5^{2}S_{1/2}$  with total angular momentum quantum number $F=2$ ($|g\rangle$) and an excited quantum state is $5^{2}P_{3/2}$ with $F=3$ ($|e\rangle$). For stationary atoms the wavelength of this transition is $\lambda\simeq780$~nm. Object laser is frequency locked to the transition and frequency is shifted by accousto-optic modulators.  Object laser frequency $\nu_{2}$ is red detuned by $-40$ MHz and frequency $\nu_{1}$ is blue detuned by $+40$ MHz from the resonant transition for stationary atoms as shown in Fig.~\ref{fig1} (b). The nearest frequency separation of object laser beams is much larger than the line width, $5.75$ MHz and much lower than Doppler broadening of the resonant transition. The frequency spread of all laser beams is less than $1$~MHz. Frequency detuning of beams is measured with a resolution $0.1$~MHz. 
An object laser beam of frequency $\nu_{o}$ is on resonance with atomic velocity class of $v_{z}=0$. Therefore, an image of an alphabet $\bf{A}$ is imprinted in the zeroth velocity class in the form of an atomic population difference. An object laser beam of frequency $\nu_{1}$ is on resonance with a velocity class $v_{z}=-31.2$~m/sec therefore, an image of an alphabet $\bf{C}$ is imprinted in this  velocity class of atoms. An object laser beam of frequency $\nu_{2}$ is on resonance with a veolcity class $v_{z}=+31.2$~m/sec therefore, an image of an alphabet $\bf{T}$ is imprinted in this velocity class of atoms.  Atoms of each velocity class are uniformly distributed in the position space volume of the atomic gaseous medium. Therefore, the imprinted pattern is completely delocalized along the length of atomic gaseous medium in the beam propagation direction. All the imprinted images form a localized pattern in an unique 3D sub-space of the 6D phase space of the atomic gaseous medium.

To image of the imprinted phase space pattern, a linearly polarized imaging laser beam is passed through the atomic medium in the opposite direction relative to the propagation direction of object laser beams. Object and imaging laser beams are produced by two independent lasers. Imaging laser is also frequency locked to the same resonant transition of stationary atoms and its frequency is shifted by accouto-optic modulators. Prior to their entrance into the atomic medium, the transverse intensity profile of imaging laser beam of frequency $\nu_{r}$ and detuning $\delta\nu=\nu_{r}-\nu_{o}$ is $I_{r}(x,y,\delta\nu)$.   Imaging laser beam is $y$-polarized which is perpendicular to the linear polarization of object laser beams and its peak intensity is much lower than the saturation intensity of the atomic transition. After passing through the atomic medium, imaging laser beam is reflected by PBS-$1$ and its transverse intensity distribution at different detunings is captured with an electron-multiplying-charge-coupled-device (EMCCD) camera without gain multiplication. Transmittance of the atomic vapour cell for the imaging laser beam, at different detuning $\delta\nu$, in presence of object laser beams without masks is shown in Fig.~\ref{fig1} (d). Three peaks labeled as $1$, $2$ and $3$ correspond to velocity selective hole-burning, in doppler broadened absorption profile, caused by object laser beams of frequencies $\nu_{1}$, $\nu_{o}$ and $\nu_{2}$, respectively. Object and imaging laser beams are counter propagating therefore, a peak in the transmittance due to a hole-burning by a higher frequency object laser beam appears at lower frequency of imaging laser beam. 
To measure the imaging laser beam detuning precisely, a part of object laser light is extracted and red detuned by $190$ MHz from the resonant transition. 
\begin{center}
\begin{figure}[h]
\begin{center}
\includegraphics[scale=0.25]{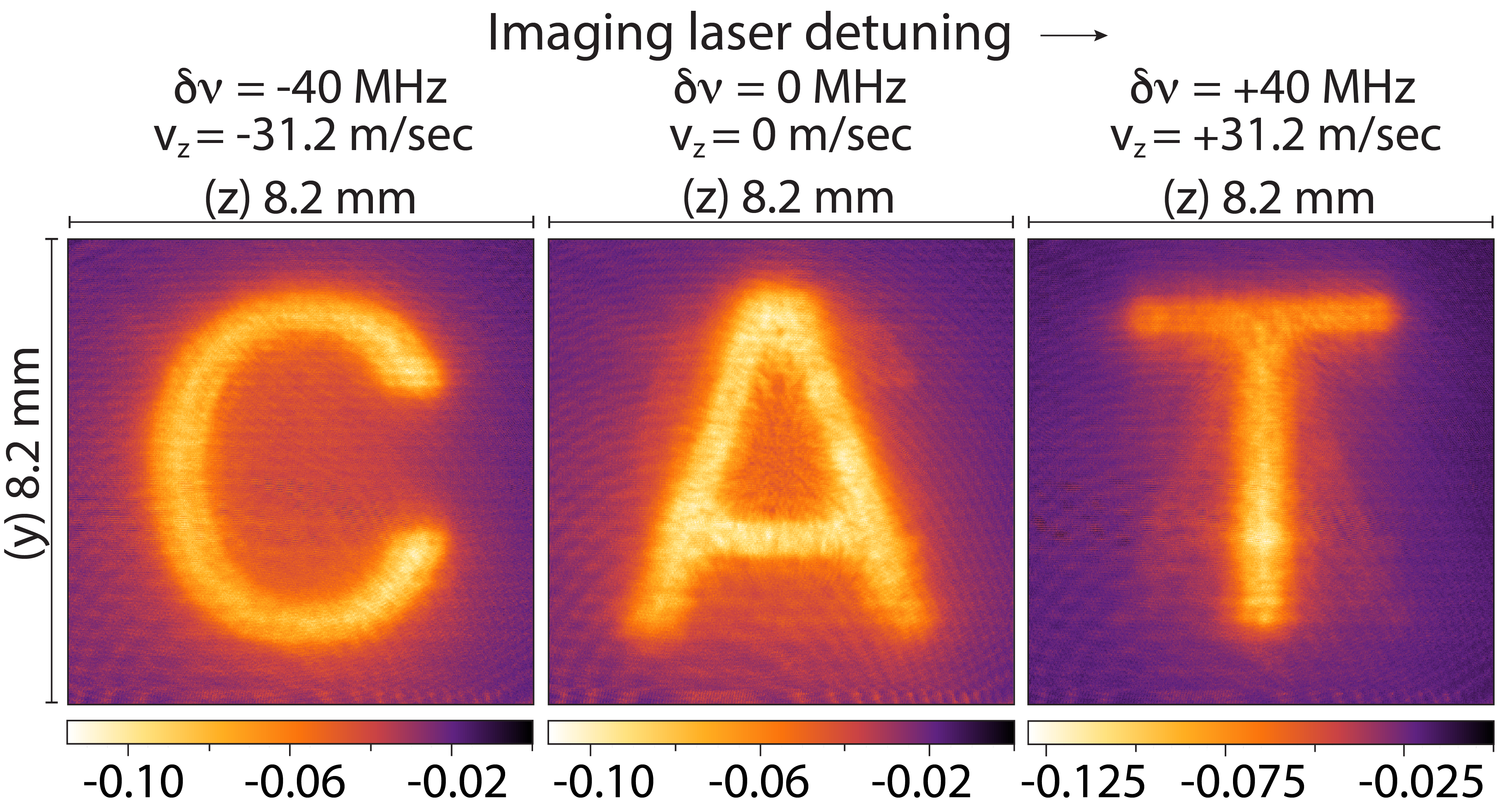}
\caption{\label{fig2} \emph{Three tomographic images, of a localized pattern in the 3D phase space, captured at different detunings of the imaging laser beam.  Each image is a plot of a change in the optical density, $\Delta \mathrm{OD}(y,z,\delta\nu)$.}}
\end{center}
\end{figure}
\end{center}
The extracted object laser light is overlapped with a part of the imaging laser light of same polarisation on a non polarization beam splitter (BS). 
A beating signal of two lasers is detected with a fast response photo detector and measured with a radio frequency spectrum analyzer as shown in Fig.~\ref{fig1} (b). Detuning is measured from frequency of the beating signal that corresponds to a frequency difference of two lasers . 

The intensity profile of imaging laser beam after traversing throught the atomic medium in presence of object laser beams is $I_{on}(x,y,\delta\nu)=I_{r}(x,y,\delta\nu) \exp(-\mathrm{OD}(x,y,\delta\nu))$. The optical density $\mathrm{OD}(x,y,\delta\nu)$ is constructed at detuning $\delta\nu$. The optical density is higher in absence of object laser beams. A change in the optical density after switching-on the object laser beams is $\Delta \mathrm{OD}(x,y,\delta\nu)=-\log(I_{on}(x,y,\delta\nu)/I_{off}(x,y,\delta\nu))$. Where, $I_{off}(x,y,\delta\nu)$ is the intensity profile of the imaging laser beam after traversing through the atomic medium in absence of object laser beams. Frequency of the imaging laser beam is red detuned by $\delta\nu=-40$~MHz from the resonant transition. Its transverse intensity profile $I_{off}(x,y,\delta\nu)$ is captured with an EMCCD camera in absence of three object laser beams.. After a time delay, another image of the imaging laser beam intensity  
$I_{on}(x,y,\delta\nu)$ is captured in presence of three object laser beams.
 A $y$-polarizer is placed in front of EMCCD camera to block any back reflection of object laser beams from optical components. A change in optical density profile, $\Delta \mathrm{OD}(x,y,\delta\nu)=-\log(I_{on}(x,y,\delta\nu)/I_{off}(x,y,\delta\nu))$, of the atomic medium is evaluated. Similar measurements are performed for detuning $\delta\nu=0$~MHz and $\delta\nu=+40$~MHz. For each detuning of the imaging laser beam, a different tomographic section of the 3D phase space localized pattern is captured. A series of three tomographic images are shown in Fig.~\ref{fig2} for three different detunings. Imaging laser beam is reflected by PBS-$1$ therefore, images are constructed after making a reflection transformation in a plane parallel to $y$-$z$ plane and $\Delta \mathrm{OD}(x,z,\delta\nu)$ is transformed to $\Delta \mathrm{OD}(y,z,\delta\nu)$. Three tomographic images resemble to the English script alphabets of object transparancies \emph{i.e.} $\bf{C}$ (at $\delta\nu=-40$~MHz), $\bf{A}$ (at $\delta\nu=0$~MHz) and $\bf{T}$ (at $\delta\nu=+40$~MHz). By combining all the tomographic images, a word $\bf{CAT}$ is formed as shown in Fig.~\ref{fig2}. Imaging laser beam detuning and $z$-component of resonating velocity class corresponding to each tomographic image are shown on top of each tomograph.
\begin{center}
\begin{figure}[t]
\begin{center}
\includegraphics[scale=0.35]{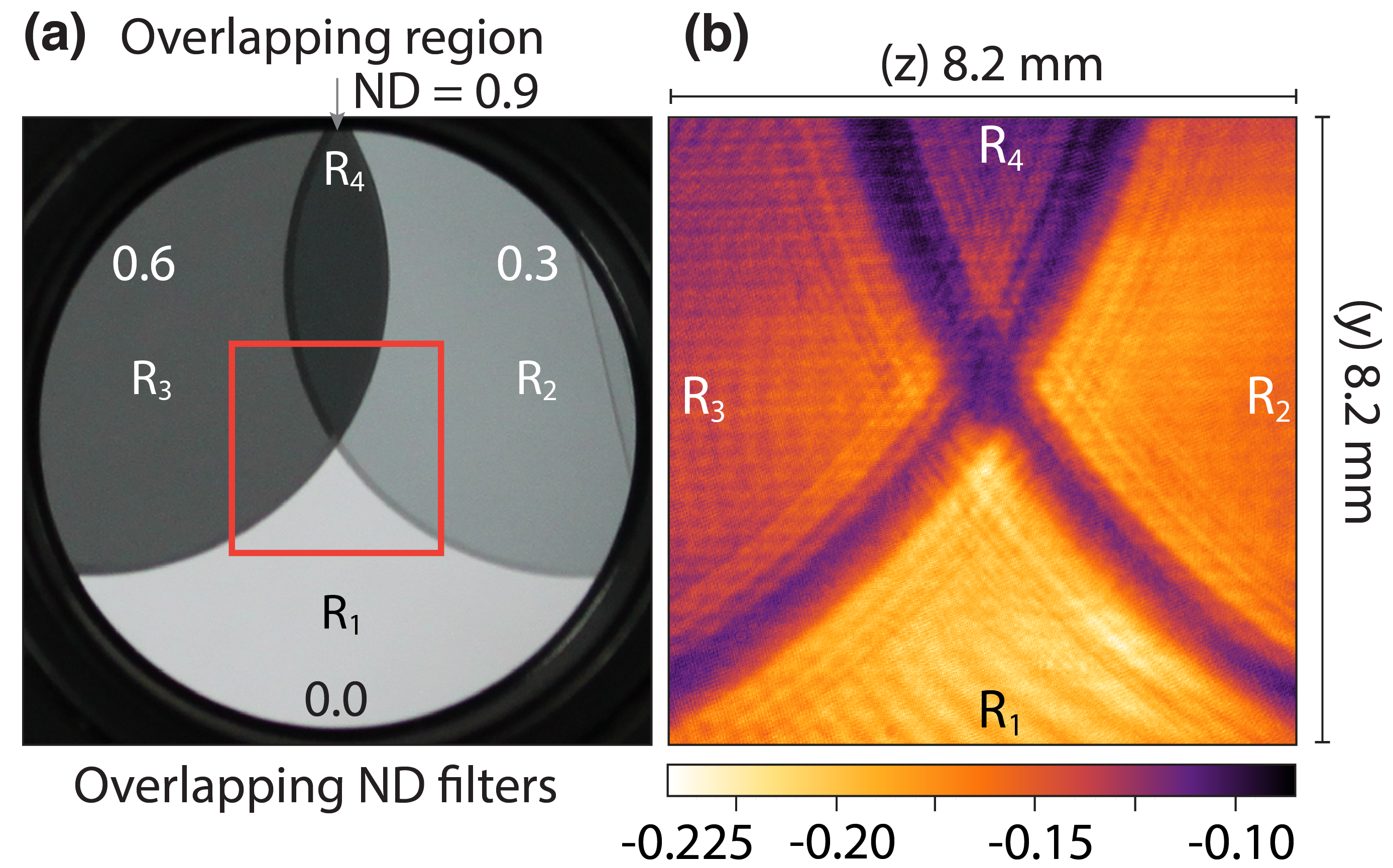}
\caption{\label{fig3} \emph{(a) A photograph of overlapping neutral density filters.  (b) An image, $\Delta \mathrm{OD}(y,z,\delta\nu=0)$, of an area enclosed by a square as shown in (a).}}
\end{center}
\end{figure}
\end{center}
In another experiment, an object of nonuniform transmittance is constructed by overlapping two neutral density filters of neutral densities ($\mathrm{ND}$) $0.3$ and $0.6$ as shown in a photograph, Fig.~\ref{fig3} (a). Four different regions $R_{1}$ ($\mathrm{ND}=0$), $R_{2}$ ($\mathrm{ND}=0.3$), $R_{3}$ ($\mathrm{ND}=0.6$) and $R_{4}$ ($\mathrm{ND}=0.9$) are formed. This object is positioned in place of a mask-$2$ in path of an object laser beam of frequency $\nu_{o}$. The image of a part of the object enclosed by a square as shown in Fig.~\ref{fig3} (a) is captured with imaging laser beam. Experiment is performed with a single object laser beam. The intensity profile $I_{p}(x,y,\delta\nu=0)$ of object laser beam after passing through the object consists of four different regions of different intensity levels. Therefore, it produces four regions of different depths of hole-burning in atomic gaseous medium. Imaging laser beam is on resonance with velocity class $v_{z}=0$ and an image of $\Delta\mathrm{OD}(y,z,\delta\nu=0)$ of the atomic gaseous medium is constructed as shown in Fig.~\ref{fig3} (b), 
which is an image of the overlapping neutral density filters. 

In conclusion, experiments presented in this paper provide a unique way to produce and image a 3D pattern localized only in an unique 3D subspace of the 6D phase space. 

\vspace{4mm}

{\bf{Contribution of Authors:} }Mandip Singh (MS) created the idea, MS designed and performed the experiment, MS and PhD student Samridhi Gambhir (SG) made the masks, SG plotted data shown in Fig. 1 (d). MS wrote the paper.

%\acknowledgments{{\bf{Acknowledgment}}:}

\end{document}